\documentclass[
              preprint,
               reprint, twocolumn,
              preprintnumbers,
               amsmath,
               amssymb,
               showpacs,
               prl,
              superscriptaddress]{revtex4-2}
              
 \usepackage{ucs}
 \usepackage[utf8x]{inputenc}
 \usepackage{latexsym}
 \usepackage{amsfonts}
 \usepackage{amsmath}
 \usepackage[dvips]{graphicx}
 \usepackage{color}
 \usepackage{bm}
 \usepackage{hyperref}
 \usepackage{lineno}
 \usepackage[autostyle]{csquotes}
\graphicspath{ {./fig/} }
\hypersetup{
   colorlinks=true,       
   linkcolor=magenta,          
   citecolor=magenta,        
   filecolor=magenta,      
   urlcolor=blue           
}

\begin{document}
\author{Krishna R. Nandipati}
\email[e-mail: ]{krishna.nandipati@pci.uni-heidelberg.de}
\affiliation{Theoretische Chemie,
             Physikalisch-Chemisches Institut,
             Universität Heidelberg,
             Im Neuenheimer Feld 229, 69120 Heidelberg, Germany}

\author{Oriol Vendrell}
\email[e-mail: ]{oriol.vendrell@uni-heidelberg.de}
\affiliation{Theoretische Chemie,
             Physikalisch-Chemisches Institut,
             Universität Heidelberg,
             Im Neuenheimer Feld 229, 69120 Heidelberg, Germany}
\affiliation{Centre for Advanced Materials,
        Universit\"at Heidelberg,
        Im Neuenheimer Feld 205,
       69120 Heidelberg, Germany}



\title{Nonadiabatic effects in electronic ring currents triggered by circularly polarized light}

\title{Electronic ring currents triggered by circularly polarized light are strongly affected by dynamical Jahn-Teller distortions.}

\title{The effect of dynamical Jahn-Teller distortions on
the generation of electronic ring currents by circularly polarized light.}

\title{Dynamical Jahn-Teller effects on
the generation of electronic ring currents by circularly polarized light}

\date{\today}


\begin{abstract}
The generation of electronic ring currents in ring-shaped molecules by
photo-excitation with circularly polarized laser light is considered in the
presence of vibronic coupling effects.
$(E\times e)$ Jahn-Teller distortions, unavoidable by symmetry in the $(E)$
subset of electronic states supporting the ring current, mix the clockwise and
anti-clockwise circulation directions of the electrons and can suppress the
maximum achievable current by at least one order of magnitude, already for
moderate vibronic coupling strengths, as compared to the Born-Oppenheimer limit
of fixed atomic positions.
%
%
The circulation direction of the electrons is found to depend on the spectral
region of the $(E\times e)$ Hamiltonian. This fact results in the surprising effect
that the same polarization direction of the laser pulse can trigger either
clockwise or anti-clockwise electronic dynamics depending on the wavelength of
the photons.
These findings are illustrated in a model of the triazine molecule.
\end{abstract}

\maketitle

There is a growing interest in inducing, probing and controlling electronic
ring-currents in ring-shaped molecular systems due to their potential
applications in the next generation of optoelectronic devices
\cite{kudisch2020ring,kaiser2019sp,saha2010quantum,barth2006unidirectional,
anthony2006functionalized,rae07:157404,ram17:303}.
Recent experiments realized ring-currents in ring-shaped organic chromophores
with strong magnetic fields
and showed that the light absorption properties of the molecules were
substantially modified~\cite{kudisch2020ring}.
From a theory perspective, it is well understood that circularly polarized light
can also trigger ring-currents by resonantly exciting the electrons to a
manifold of doubly-degenerate electronic states of $(E)$ symmetry related to a
symmetry axis of the molecule~\cite{barth2006unidirectional,barth2010quantum,%
yuan2017attosecond,jia2017quantum,kanno2018laser,liu2018attosecond}.
%
It is inevitable by symmetry, though, that the $(E\times e)$ Jahn-Teller (JT)
effect couples the $(E)$ electronic states and $(e)$ vibrational modes
in all molecular point groups able to support ring
currents~\cite{jahn1937stability,Longuet-Higgins,englman,koppel},
thus potentially mixing the two pristine ring-current circulation directions
defined in the uncoupled
limit~\cite{nandipati2020generation,kanno2010nonadiabatic}.
The role of the $(E\times e)$ JT effect and related vibrational-electronic
(vibronic) coupling terms is
well established in the areas of electronic
spectroscopy~\cite{domcke2004conical,domcke2011conical}, ultrafast vibronic
dynamics~\cite{li13:038302,arn17:33425} and charge migration~\cite{des15:426}.
However, it is not yet clear how vibronic couplings effects, particularly
the dynamical $(E\times e)$ JT effect~\cite{Longuet-Higgins}, modify
the properties of the ring-currents, limit their magnitude and stability, and
thus their controllability.

In this letter, we describe the theory of photo-induced molecular ring currents
under nuclear-electronic couplings. We then apply this theory to the general $(E
\times e)$ JT Hamiltonian, where the vibronic coupling results in
a conical intersection (CI) between the corresponding adiabatic potential energy
surfaces and to a complex vibronic spectrum at moderate to strong
couplings~\cite{Longuet-Higgins}.
The $(E \times e)$ JT Hamiltonian constitutes the basis on which the description of
more complex vibronic interactions, involving multimode effects in polyatomic
molecules, have been built~\cite{koppel}. Its
general features have been well known since the
pioneering works of Longuet-Higgins and
others~\cite{moffitt1957configurational,Longuet-Higgins,sturge,englman}, making it
an ideal model to unravel the connection between molecular ring currents and
non-adiabatic effects in molecules.
The dramatic effects of the vibronic coupling on the photo-induced ring-currents
are finally demonstrated on the JT Hamiltonian of the sym-triazine
molecule~\cite{whetten1986dynamic}.

Within a diabatic representation of electronic states $|\Phi_l(\bm{Q}_0)\rangle$
defined at the reference nuclear configuration $\mathbf{Q}_0$~\cite{koppel}, the
$j$-th eigenstate of the total molecular Hamiltonian
can be written as
  \begin{equation}\label{BH}
      |\Psi_j\rangle = \sum_{l}
      |\Xi_{l}^{(j)}\rangle \otimes |\Phi_l(\bm{Q}_0)\rangle,
  \end{equation}
where $|\Xi_{l}^{(j)}\rangle$ indicates the nuclear wavefunction corresponding
to the $l$-th electronic state. Here and in the following, bold face quantities
$\bm{q}$ and $\bm{Q}$ denote the collective coordinates of electrons and nuclei,
respectively, and the $\vec{q}_r=(x,y,z)_r$ indicate the spatial coordinates of
the $r$-th electron.

We consider now the doubly degenerate ($E$) subspace of excited electronic
states ($|\Phi_{x}(\bm{Q}_0)\rangle$, $|\Phi_{y}(\bm{Q}_0)\rangle$) defined at
the minimum energy geometry, $\bm{Q}_0$,
of the totally symmetric ground electronic state
$|\Phi_{A}(\bm{Q}_0)\rangle$. Doubly degenerate subspaces are always present in
molecules with a rotation axis $C_n$ of order $n\geq 3$, which enable degenerate
electronic ring currents in the two rotation directions around this
symmetry element
\cite{barth2010quantum,barth2006unidirectional,barth2006periodic}.
To define these currents, one needs to introduce the complex linear combination
of the two real orthonormal electronic configurations
\begin{equation}\label{delbasis}
    |\Phi_{\pm}(\bm{Q_0})\rangle = \frac{1}{\sqrt{2}}
      \big(
        |\Phi_{x}(\bm{Q_0})\rangle \pm i |\Phi_{y}(\bm{Q_0})\rangle
      \big).
\end{equation}
The vibronic coupling (cf. Eq.~(\ref{HVpolar})) mixes the two $(E)$ electronic
configurations and the vibronic eigenstates $|j\rangle$
of the total molecular Hamiltonian
can thus each be written as
  \begin{align}
      \label{jth}
      |j\rangle & =
      |\Xi_{+}^{(j)}\rangle \otimes |\Phi_+(\bm{Q}_0)\rangle \\\nonumber
      & +
      |\Xi_{-}^{(j)}\rangle \otimes |\Phi_-(\bm{Q}_0)\rangle.
  \end{align}
  In the above expression, we assume that the vibronic coupling mechanism only
  mixes the two components of the ($E$) states with each other, which is the
  case in the $(E \times e)$ JT Hamiltonian. In general though, the ($E$) states
  may also couple to states of other symmetry representations, for example ($A$)
  and/or ($B$), in particular point
  groups~\cite{koppel,domcke2004conical,domcke2011conical}.  These other
  electronic states do not contribute to the ring current and we can neglect
  them for the purposes of our analysis.
  Starting with Ansatz~(\ref{jth}), and inserting the electronic current
  operator related to the electronic charge density via the continuity
  equation~\cite{Messiah1999}, one arrives at an expression for the one-body
  electronic current of the $j$-th \emph{vibronic eigenstate},
  $\vec{J}_j(\vec{q})$ (cf. the supporting information (SI)~\cite{SI} for the complete derivation).
  By introducing cylindrical coordinates $\vec{q}=(r,\theta,z)$ for the
  electrons, integrating $\vec{J}_j(\vec{q})$ over the radial and axial
  coordinates, and averaging the remaining flux over the $\theta$ angle, one
  arrives at~\cite{SI}
\begin{align}\label{JR}
    \mathcal{C}_{j}
    & = \frac{\hbar}{m_e} \bigg(P_{+}^{(j)} - P_{-}^{(j)}\bigg)
\end{align}
   for the \emph{ring} current of the vibronic eigenstate
   $|j\rangle$ of the full vibrational-electronic Hamiltonian,
   where $P_{\pm}^{(j)}=\langle \Xi_{\pm}^{(j)} | \Xi_{\pm}^{(j)} \rangle$.
This state-specific current is proportional to the imbalance
of the population of the $E_+$ and $E_-$ electronic states in
$|\Psi_j\rangle$.
If one considers the time-dependent Born-Huang
expansion~\cite{born1954dynamical}
  \begin{align}
      \label{td}
      |\Psi (t)\rangle =
      |\Xi_{+}(t)\rangle \otimes |\Phi_+(\bm{Q}_0)\rangle +
      |\Xi_{-}(t)\rangle \otimes |\Phi_-(\bm{Q}_0)\rangle
  \end{align}
instead of the eigenstates~(\ref{jth}), the same train of arguments follows
and the time-dependent current of the wavepacket reads~\cite{SI}
\begin{align}\label{JRtd}
    \mathcal{C}(t)
    = \frac{\hbar}{m_e} \bigg(P_{+}(t) - P_{-}(t)\bigg).
\end{align}
Equations~(\ref{JR}) and (\ref{JRtd}) are our main working equations.

%
%
All molecular point groups with at least one rotational axis of order $n\geq3$
have one ($E$)-representation that transforms like the $(x,y)$ components of the
dipole operator, where $(x,y)$ is the plane of rotation perpendicular to the
axis~\cite{barth2008quantum,englman}. These are, for example, the ($E_{1u}$) and ($E_u$)
representations in the $D_{6h}$ and $D_{4h}$ point groups of benzene~\cite{liu2018attosecond} and
porphyrins~\cite{barth2006periodic}, respectively.
Therefore, one-photon electronic transitions from the totally symmetric
ground-electronic state to the $(E)$ states are allowed and the interaction of
the molecules with circularly polarized light in the molecular $(x,y)$-plane can
generate ring currents owing to angular momentum conservation in the
photo-absorption process \cite{eckart2018ultrafast,nandipati2020generation}.

The interaction between the molecule and light propagating along the
$z$-direction is described in the electric dipole approximation and the
electromagnetic radiation is treated classically.
The interaction term thus reads
\mbox{$ - (\hat{\mu}_x \mathcal{E}_x(t) + \hat{\mu}_y \mathcal{E}_y(t))$},
with transition dipole operators
$\hat{\mu}_v = \mu_{AE}(|\Phi_v(\bm{Q}_0)\rangle \langle\Phi_A(\bm{Q}_0)| +
h.c.)$.
The electric field $\mathcal{E}_v(t)$ is derived from the vector potential
$\mathcal{A}_v(t) = \frac{\mathcal{E}_0}{\omega_L}S(t)\sin({\omega_L t -\phi_v})$
as $E_v(t)=-\partial A_v(t) / \partial t$. The pulse envelope
is taken to be \mbox{$S(t) = \tilde{\Theta}(t-\tau)\sin^2 \left(
\frac{\pi t}{\tau} \right)$}, and
$\mathcal{E}_0$, $\tau$ and $\omega_L$ are the maximum amplitude, pulse duration
(start to end) and carrier frequency of the pulse, respectively, and
$\tilde{\Theta}(t-\tau)$ is the inverse Heaviside step function.
 For circular polarization, all pulse parameters for both $v\to(x,y)$
 polarization directions are taken to be equal except for the $\phi_v$ phases.
 These are $(\phi_x=0, \phi_y=\pi/2)$ for a left circularly polarized pulse
 (LCP) and $(\phi_x=\pi/2, \phi_y=0)$ a right circularly polarized pulse (RCP).

%
%
We apply the above considerations to the $(E \times e)$ JT model system with 
Hamiltonian
\begin{align}\label{HVpolar}
    \hat{H} = & \hat{T}_N + \hat{H}_{el} \\\nonumber
            = & \hat{T}_N +
 \frac{\omega}{2} \rho^2 \bm{I}_{3\times 3}
  +
  \left(
    {\begin{array}{ccc}
            \epsilon_+   &   0  & \kappa \rho e^{-i\alpha} \\
          0 &  \epsilon_A            &  0        \\
         \kappa \rho e^{i\alpha}  &  0    & \epsilon_-
    \end{array}}
\right)
\end{align}
written here in its polar representation for the $(e)$ vibrational modes
$(\rho,\alpha)$ \cite{Longuet-Higgins,koppel}.
The electronic part of the total molecular
Hamiltonian, $\hat{H}_{el}$, is represented in the diabatic
basis $\{ |\Phi_{A}(\bm{Q}_0)\rangle ,|\Phi_{\pm}(\bm{Q}_0)\rangle\}$ and
includes off-diagonal
first-order JT interaction between the $(E)$ electronic states
and $(e)$ vibrational modes~\cite{koppel}.
\textbf{\textit{I}}$_{3 \times 3}$ is the 3 $\times$ 3 unit matrix,
$\omega$, $\epsilon_{\pm,A}$ and $\kappa$ are the
frequency of the $(e)$ modes, the energy of the diabatic
electronic states at the reference geometry $\bm{Q}_0$ (i.e. $\rho=0$),
and the linear
JT coupling parameter, respectively.
The vibronic interaction between the two $(E)$ states can be characterized by the
dimensionless coupling strength parameter ${\kappa}/{\omega}$.
The diagonalization of the $H_{el}$ matrix as a function of the
$(\rho,\alpha)$ coordinates yields the adiabatic potential energy surfaces
with the ``Mexican-hat'' shape meeting at a CI
at $\rho=0$~\cite{bersuker2006jahn}.

%
%
The effect of the vibronic coupling on the ring currents is most conveniently
described by introducing the eigenstates of the 2D-harmonic oscillator (HO) in
the polar representation
$\chi_{n,m}(\rho,\alpha) = \langle \rho,\alpha|n,m\rangle$,
where $n$ and  $m$ are the radial and angular momentum
quantum numbers of the 2D-HO~\cite{paulingandwilson}.
These can be extended to $|n,m,l\rangle$, including the $l$ quantum number for
the electronic angular momentum operator around the rotational axis and defined by
$\hat{L}_{el}|\Phi_\pm(\bm{Q}_0)\rangle = \pm\hbar\; |\Phi_\pm(\bm{Q}_0)\rangle$ and
$\hat{L}_{el}|\Phi_A(\bm{Q}_0)\rangle = 0$.
%
The total angular momentum operator in the $(E\times e)$ JT system thus reads
$-i\hbar 2\partial/\partial\alpha + \hat{L}_{el}$
(cf. Ref.~\cite{Longuet-Higgins} for a discussion on the factor 2 for the
vibrational term) and a \emph{ring} current operator
$\hat{\mathcal{C}} = \hat{L}_{el}/m_e$  can be introduced by comparing
Eqs.~(\ref{JR},\ref{JRtd}) with the definition of $\hat{L}_{el}$.
Now, the vibronic eigenstates of the total Hamiltonian can be classified by the
vibronic angular momentum quantum number $q=2m+l$, which is conserved under the
vibronic coupling~\cite{Longuet-Higgins}.
The totally symmetric $|\Phi_A(\bm{Q}_0)\rangle$ electronic state participates
only in the $q=0$ block, whereas the $|\Phi_\pm(\bm{Q}_0)\rangle$ states span
both $q=\pm1$ blocks due to the vibronic coupling. The JT coupling matrix
elements of $\hat{H}$ in the $|n,m,l\rangle$ basis are given in the supporting
information for completeness~\cite{SI}.

%
Dipolar transitions from the ground electronic state fulfill the selection rule
$\Delta q=\pm1$ and circularly polarized radiation can only induce transitions
$(0 \leftrightarrow 1)$ for LCP and $(0 \leftrightarrow -1)$ for
RCP~\cite{nandipati2020generation}.
Finally, the vibronic eigenstates $|j_q\rangle$ of the complete
Hamiltonian~(\ref{HVpolar}) in the $q=1$ block are one-to-one degenerate with
the eigenstates of the $q=-1$ block. The degenerate pairs are related by the
reversal of the signs of the $l$ and $m$ quantum numbers in all basis states
$|n,m,l\rangle$ in their basis set
expansion~\cite{Longuet-Higgins,whetten1986dynamic}.
As a result, the electronic populations and currents of the vibronic eigenstates
fulfill $P_\pm^{(j_{\pm 1})} = P_\mp^{(j_{\mp 1})}$ and $\mathcal{C}_{j_{+1}} =
-\mathcal{C}_{j_{-1}}$ (cf. Eq.~(\ref{JR}) and \cite{SI}).

\begin{figure}[t]
  \centering
 \includegraphics[width=0.95\columnwidth]{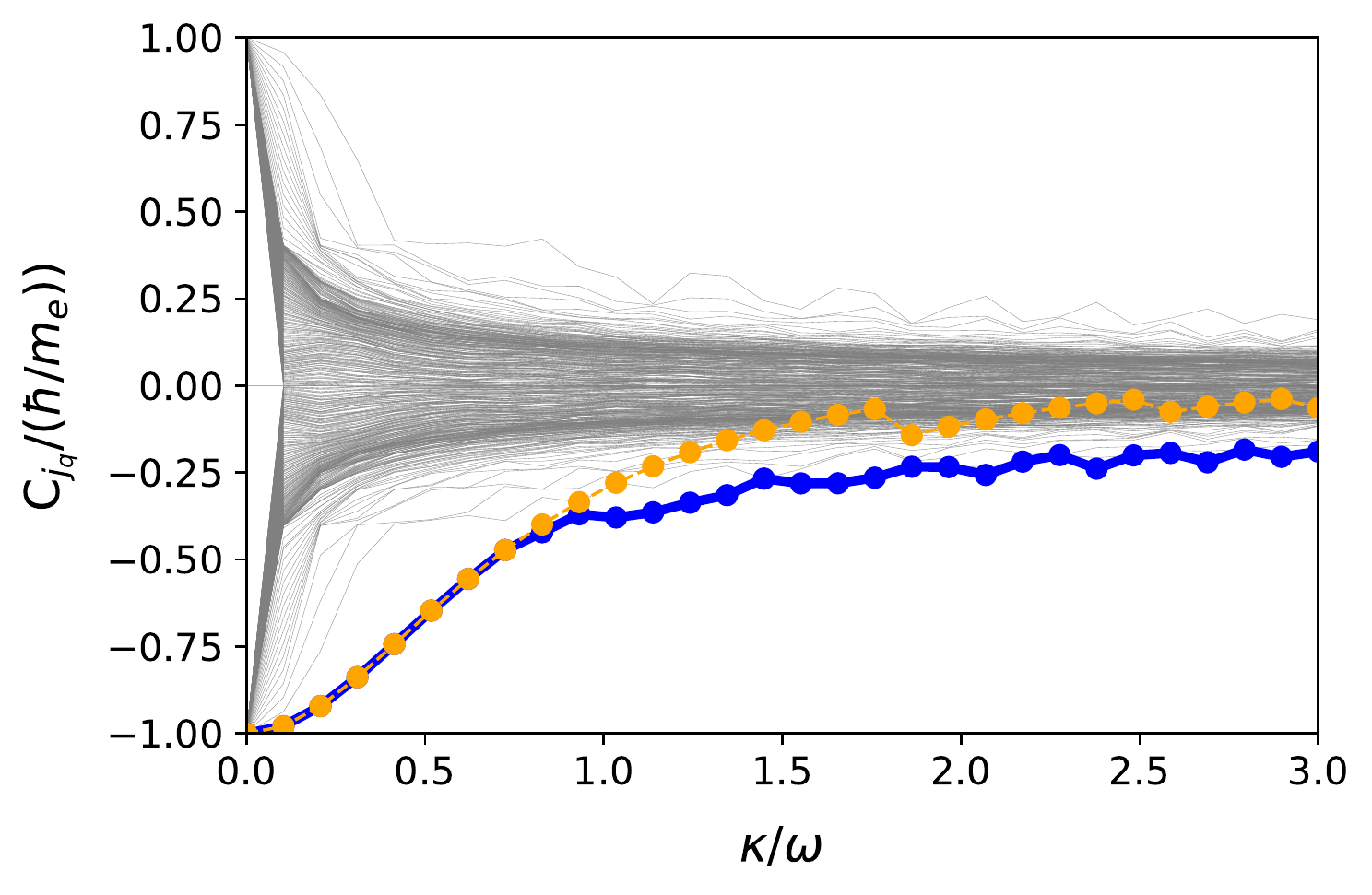}
  \caption{
  \label{currents_kappa}
    Ring current ${C}_{j_q}$ values in $\hbar/m_e$ units
    for the eigenstates of the $q=-1$ block as a
    function of the coupling strength $\kappa/\omega$. The dot-dashed curve in
    orange represents the current of the most optically bright eigenstate while
    the blue curve indicates the largest ring current achievable at a specific
    coupling strength.
    }
\end{figure}

%
%
In the following discussion we fix $\omega=0.003$~au ($\approx 660$~cm$^{-1}$),
and $\epsilon_\pm - \epsilon_A = 7$~eV, both in the typical range for molecular
vibrations and vertical electronic transitions in molecules as, e.g., sym-triazine ($D_{3h}$)\cite{whetten1986dynamic} and benzene ($D_{6h}$) \cite{worth2007model}.
The magnitude of the ring current
$\langle j_q | \hat{\mathcal{C}} | j_q\rangle m_e/\hbar$
(in dimensionless units)
is strongly affected by the vibronic coupling strength
$\kappa/\omega$, as seen in Fig.~\ref{currents_kappa} for the $q=-1$ block.
In the weak vibronic coupling regime, i.e. ${\kappa}/{\omega}<<1$, the ring
current supported by each eigenstate is close to either 1 or -1.
In the limit $\kappa\to0$, $|0,0,-1\rangle$ coincides with the eigenstate with
the largest transition dipole matrix element $|\langle
0,0,0|\hat{\mu}|j_{-1}\rangle|$, which is also the state that supports the
largest ring current (in absolute terms).
As the vibronic coupling increases towards $\kappa/\omega=1.0$, the state with
the largest transition dipole
does not necessarily
coincide with the state featuring the largest current
(cf. blue and orange curves in Fig.~\ref{currents_kappa}).
The state-dependent ring current of all states quickly decreases as the coupling
increases due to the progressive mixing of the left ($E_{-}$) and right
($E_{+}$) circulations of the electrons and the participation of high-lying
vibrational levels.
For stronger coupling, ${\kappa}/{\omega}>2$, the largest attainable current for a single eigenstate
reaches an average $\sim$20\% of the magnitude of the current achievable in the zero coupling
limit (or by artificially fixing the molecular geometry to $\bm{Q}_0$).

%
%
 \begin{figure*}[t]
  \centering
  \includegraphics[width=8.0cm]{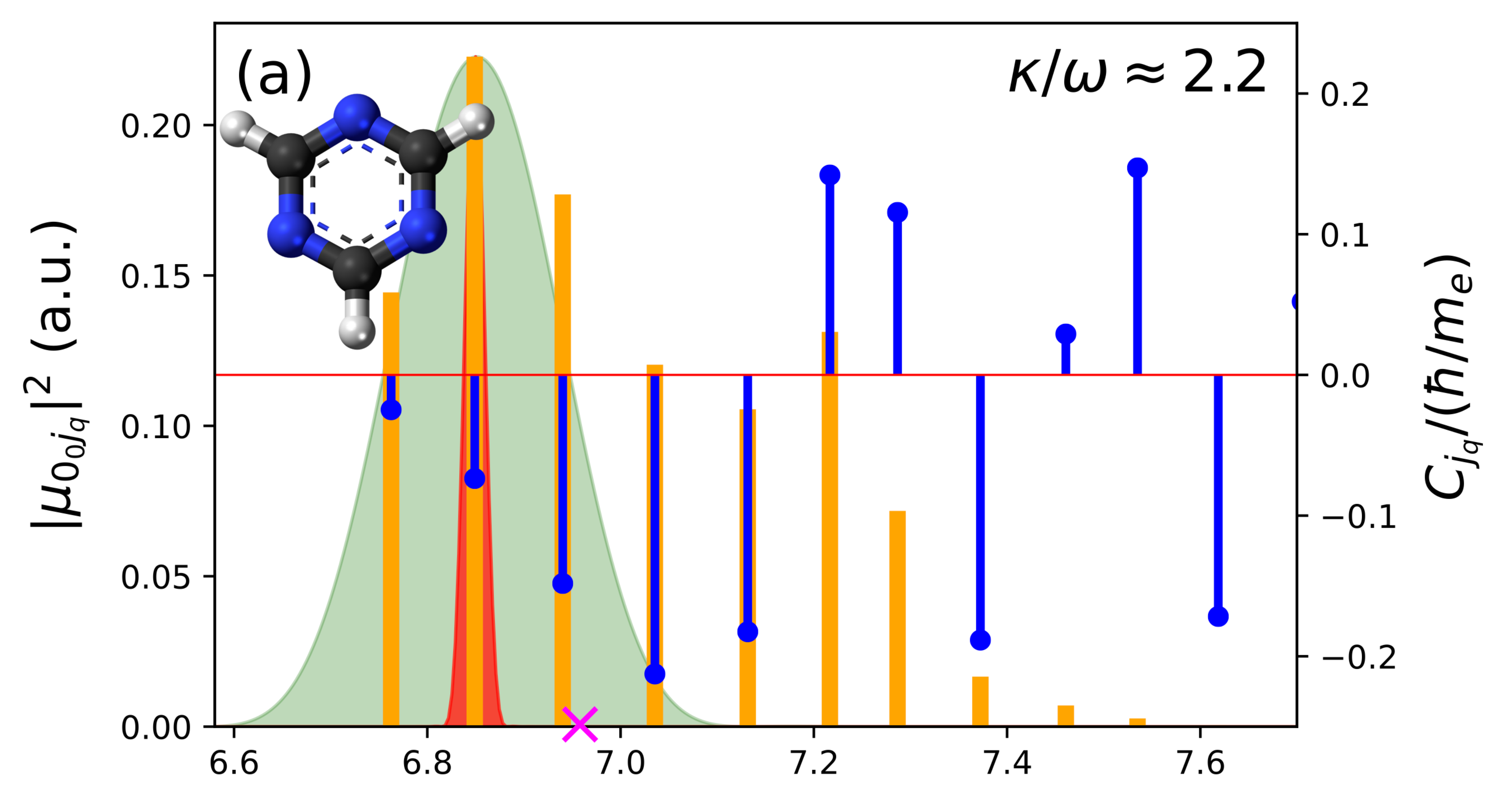}
  \hspace{-0.3cm}
  \includegraphics[width=6.8cm]{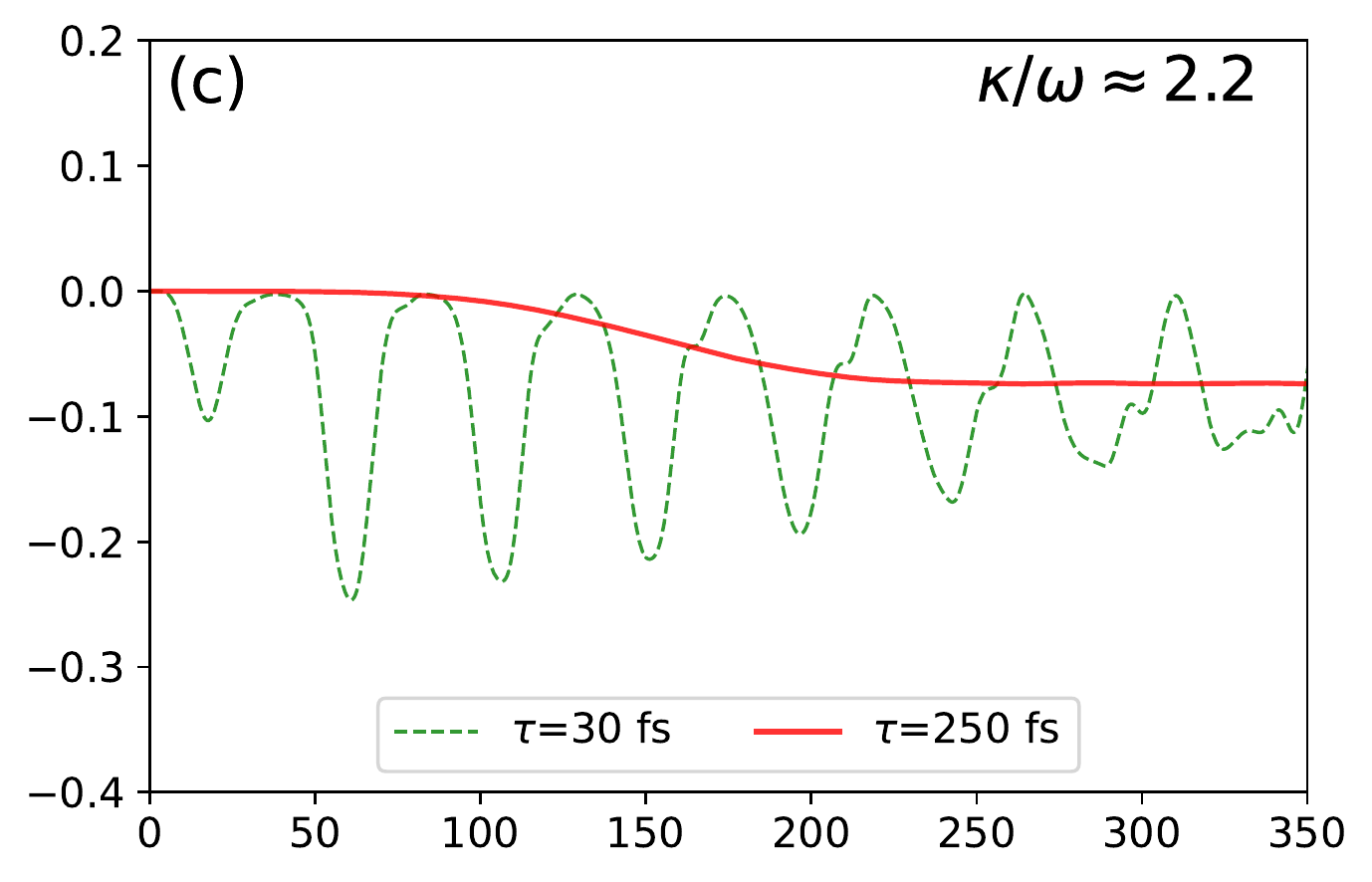}
  \includegraphics[width=8.0cm]{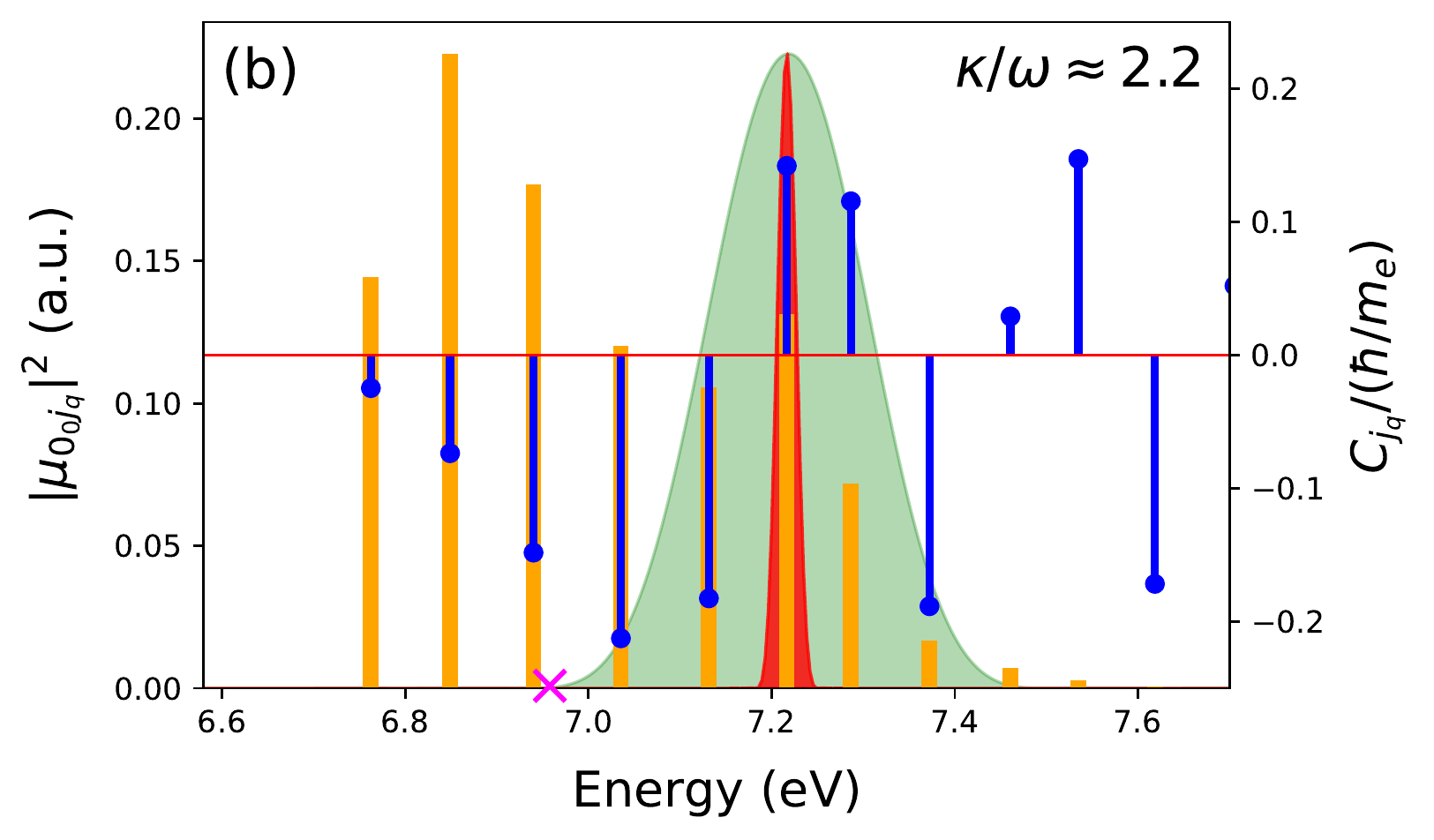}
  \hspace{-0.3cm}
  \includegraphics[width=6.8cm]{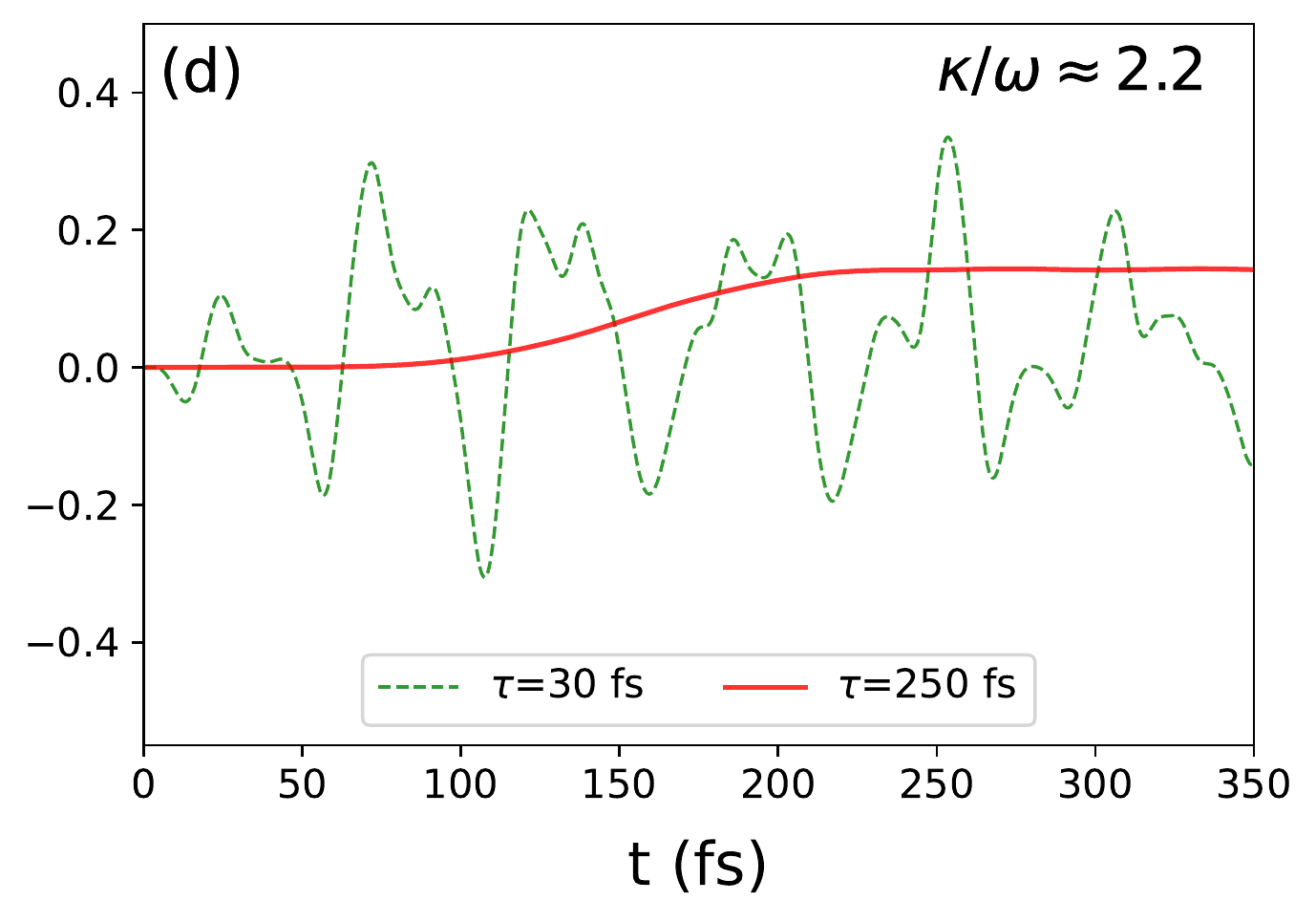}
  \caption{
  \label{currents_kappa_gen}
     (a) Absorption spectrum ($\propto|\mu_{0_0j_q}|^2$) for the 2D sym-triazine $(E \times e)$ JT Hamiltonian
     (orange, left y-scale) and eigenstate ring currents ($\mathcal{C}_{j_q}/(\hbar/m_e)$)
     (blue, middle y-scale) as a function of the eigenstates' energy superimposed
     with the spectral representation of the short
     ($\tau$=30 fs, shaded in green) and long ($\tau$=250 fs, shaded in red)
     LCPs resonant with the lower band
     of the absorption spectrum.
     (b) Same as (a) but pulses are
     resonant with the spectral region above the CI
     (marked \textcolor{magenta}{$\times$} on the energy axis).
     (c, d) Time-dependent ring currents ($\mathcal{C}(t)/(\hbar/m_e)/P_{ex}(t_f)$)
     generated
     by the pulses shown in (a, b), respectively.
     The time-dependent ring currents are
     normalized by the total excitation probability $P_{ex}(t_f)$ after the
     pulse (cf. main text). The inset in (a) shows the geometry of sym-triazine 
     and the horizontal red line in (a) and (b) is to guide the eye.} 
     %
 \end{figure*}
  The stationary ring current in a specific molecular eigenstate (cf.
  Eq.~(\ref{JR})) can be achieved by a long and narrow-bandwidth circularly
  polarized pulse, whereas time-dependent ring currents (cf. Eq~(\ref{JRtd}))
  can be obtained with a circularly polarized pulse of sufficient bandwidth that
  is resonant with a set of molecular eigenstates of the same $q$-block.
  While considering photo-triggered ring currents, we set the product of the electric
  field amplitude and the transition dipole matrix element,
  $\mathcal{E}_0\cdot\mu_{AE}$, low enough to remain in the regime of
  first-order perturbations. In this regime, the ring currents are
  intensity-independent when normalized by the amount of total excitation
  $P_{ex}(t_f)=(1-P_0(t_f))$, where $P_0(t)$ is the population of the absolute ground state
  and $t_f$ is the final time of the simulation after the pulse is over.
  $P_{ex}(t_f)$ is kept in the order of 5$\times 10^{-2}$ or lower.
  %



  In the following we consider photo-triggered ring currents in an $(E\times e)$
  JT system with a $\kappa/\omega\approx 2.2$, the value corresponding to
  the sym-triazine system \cite{whetten1986dynamic}. 
  We consider both long (250~fs) and short (30~fs) RCP pulses,
  whose spectrum is shown superimposed in Figs.~\ref{currents_kappa_gen}a-b in
  red and green, respectively.
  At about 6.8~eV, the long pulse is resonant with the eigenstate featuring the
  largest transition dipole moment.
  This state is part of the lower energy absorption band, which is
  energetically located below the CI (marked with
  a cross on the energy axis).
  The bandwidth of the pulse is narrow enough to overlap effectively with only
  one vibronic eigenstate, which results in a stationary ring current after the
  pulse is over, as indicated by the red curve
  in Fig.~\ref{currents_kappa_gen}c. As expected, the excitation-normalized
  stationary ring current coincides with the ring current of the corresponding
  eigenstate: $\mathcal{C}_{j_q} = \mathcal{C}(t_f)/P_{ex}(t_f)$, about -0.075 in
  $\hbar/m_e$ units.
  The shorter pulse, in contrast, overlaps with several eigenstates, resulting
  in a coherent excitation and an oscillatory ring current (cf. the green curve in  Fig.~\ref{currents_kappa_gen}c). Because the eigenstates in this spectral region are characterized by an clockwise electronic
  circulation (as seen from the laser source), as indicated by the negative $C_j/(\hbar/m_e)$ values
  shown in blue in Fig.~\ref{currents_kappa_gen}a, the
  coherent oscillatory current does not change its direction and its time-averaged value is negative.

  As a comparison, we consider now the spectral region at about 7.2~eV, and thus
  above the CI as shown in Fig.~\ref{currents_kappa_gen}b. Specific eigenstates in this spectral
  region can be identified with either right- or left-circulation of the electrons.
  The long pulse is resonant with one single eigenstate and results, again,
  in a stationary ring current (cf. red curve in Fig.~\ref{currents_kappa_gen}d), although of larger
  magnitude ($\sim$0.14) than before and, remarkably, with opposite circulation direction
  of the electrons as compared to the long pulse with a lower photon energy.
  Thus, different spectral regions of a JT system may feature
  opposite circulation directions of the electrons, which is a direct
  consequence of
  the vibronic coupling effects and of the mixing of the electronic and
  vibrational angular momenta. Therefore, a RCP does not necessarily result in
  clockwise
  circulation of the electrons, as in the $\kappa\to0$ limit.
  A shorter pulse in this spectral region results in an oscillatory current (cf. green curve in Fig.~\ref{currents_kappa_gen}d) which, however, oscillates now with both positive and negative ring current values. The ring current changes direction periodically although it remains positive when time-averaged.

  Finally, it is interesting to note that the energetic proximity of the
  CI at about 7~eV has no particular consequences for the ring currents. The
  ground vibronic eigenstate $|0_{-1}\rangle$ of the $q=-1$ block is found to be at about 6.8~eV
 with the current approaching zero. This state lies further deep in the ``Mexican-hat'' adiabatic potentials at the coupling  ${\kappa}/{\omega} >$ 2.5 (not shown), where the mixing of both
  circulation directions is very strong and no currents can be generated in this
  spectral region. As the
  eigenstates' energy increases and approaches the CI energy, $\mathcal{C}_{j_q}$ and
  $|\mu_{0_0 j_q}|^2$ keep varying smoothly and no
  special effects related to the CI are present.
  Nonetheless, even at relatively large
  vibronic couplings it is possible to identify spectral regions where either
  stationary or oscillatory ring currents can be generated.
  Due to the vibronic coupling, however,
  these currents are
  about one order of magnitude smaller than the currents that can be generated
  in the $\kappa\to0$ limit (in which case $\mathcal{C}$/($\hbar/m_e$) is $\pm$ 1). We remark that these and the above claims are not affected by the inclusion of second-order JT coupling [cf. Fig.1 in SI and its discussion]. 

  In conclusion, molecular systems with the necessary symmetry in their ground
  electronic state support stationary and oscillatory ring currents upon
  photo-excitation to an $(E)$-representation manifold of electronic states by
  circularly polarized light, even in the presence of strong vibronic coupling
  effects. We have arrived at expressions for such currents in terms of vibronic
  eigenstates and of time-dependent nuclear-electronic wavepackets, and applied
  the theory to the paradigmatic $(E\times e)$ JT Hamiltonian.
  The observations made on JT systems are general and have consequences for
  laser-generated ring currents and their control in complex
  molecules, and indicate that vibronic coupling effects are central and must be considered
  in future applications.

\begin{acknowledgments}
    K.N. acknowledges the collaborative research center ``SFB 1249: N-Heteropolyzyklen als Funktionsmaterialen''
    of the German Research Foundation (DFG) for financial support. The authors thank Prof. Horst Köppel and Prof. Wolfgang Domcke for helpful discussions.
\end{acknowledgments}

\bibliography{ecurrents}
\end{document}